\documentclass[floatfix,aps,pra,twocolumn,a4paper,superscriptaddress,nofootinbib,balancelastpage]{revtex4-1}
\usepackage[hidelinks,breaklinks=true]{hyperref}
\usepackage[english]{babel}
\usepackage{amsopn,amsthm,dsfont}
\DeclareMathOperator{\Tr}{Tr}
\newcommand{\ket}[1]{|#1\rangle}
\newcommand{\bra}[1]{\langle #1|}
\newcommand{\braket}[2]{\langle #1 | #2 \rangle}
\usepackage{mathtools}
\mathtoolsset{showonlyrefs=true}
\usepackage{tikz}
\usetikzlibrary{positioning}
\usepackage{adjustbox}
\usepackage[caption=false]{subfig}
\newtheorem{theorem}{Theorem}
\newtheorem{corollary}{Corollary}
\newtheorem{lemma}{Lemma}
\newtheorem{definition}{Definition}
\newtheorem{example}{Example}

\begin{document}
\title{The space of logically consistent classical processes without causal order}
\author{\"Amin Baumeler}
\author{Stefan Wolf}
\affiliation{Faculty of Informatics, Universit\`{a} della Svizzera italiana, Via G.\ Buffi 13, 6900 Lugano, Switzerland}
\affiliation{Facolt\`{a} indipendente di Gandria, Lunga scala, 6978 Gandria, Switzerland}
\begin{abstract}
	\noindent
	Classical correlations without predefined causal order arise from processes where parties manipulate random variables, and where the order of these interactions is not predefined.
	No assumption on the causal order of the parties is made, but the processes are restricted to be logically consistent under any choice of the parties' operations.
	It is known that for three parties or more, this set of processes is larger than the set of processes achievable in a predefined ordering of the parties.
	Here, we model all classical processes without predefined causal order geometrically and find that the set of such processes forms a polytope.
	Additionally, we model a smaller polytope --- the deterministic-extrema polytope --- where all extremal points represent deterministic processes.
	This polytope excludes probabilistic processes that must be --- quite unnaturally --- {\em fine-tuned}, because any variation of the weights in a decomposition into deterministic processes leads to a logical inconsistency.
\end{abstract}

\maketitle

\section{Motivation and main result}
\begin{figure}
	\centering
	\begin{tikzpicture}
		\node[draw,shape=rectangle,minimum width=2cm,minimum height=1cm] (S) {};
		\node[right=0.3cm of S.west,inner sep=0pt,outer sep=0pt] (St) {$X$};
		\node[left=0.3cm of S.east,draw,shape=rectangle,minimum width=0.8cm,minimum height=0.3cm] (M) {};
		\draw (M.east) arc (70:110:1.18);
		\draw[-] (M.south)++(0cm,0.05cm) -- ++(0.3cm,0.2cm);
		\node[draw,shape=rectangle,minimum width=2cm,minimum height=1cm,right=of S,inner sep=0pt,outer sep=0pt] (T) {};
		\node[right=0.3cm of T.west,inner sep=0pt,outer sep=0pt] (Tt) {$A$};
		\node[left=0.4cm of T.east,draw,shape=circle,inner sep=0pt,outer sep=0pt,minimum size=0.1cm] (K) {};
		\node[draw,shape=circle,inner sep=0pt,outer sep=0pt,minimum size=0.4cm] (K2) at (K) {};
		\foreach \angle in {
			-13+0*30,
			-13+1*30,
			-13+2*30,
			-13+3*30,
			-13+4*30,
			-13+5*30,
			-13+6*30,
			-13+7*30,
			-13+8*30,
			-13+9*30,
			-13+10*30,
			-13+11*30,
			-13+12*30
		}
		{
			\draw (K.center)++(\angle:0.05cm) -- +(\angle:0.15cm);
		}
		\draw[->] (T.west) -- (S.east);
	\end{tikzpicture}
	\caption{If the random variable~$A$ is an input (here, visualized by a knob), the random variable~$X$ is an output, and~$A$ is correlated to~$X$, then~$A$ can signal to~$X$ which implies that~$X$ is in the {\em causal future\/} of~$A$ ($X\succeq A$).}
	\label{fig:causality}
\end{figure}
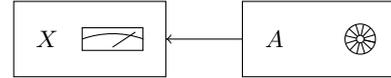

An assumption often made in physical theories, sometimes implicitly, is the existence of a {\em global time}.
In particular, quantum theory is formulated with time as an intrinsic parameter.
If one relaxes this assumption by requiring {\em local\/} validity of some theory and {\em logical\/} consistency only, then a larger set of correlations can be obtained, called {\em correlations without predefined causal order}.
The processes that lead to such correlations are called processes without predefined causal order.
Two motivations to study such correlations are quantum gravity and quantum non-locality.
Quantum gravity motivates this research in the sense that on the one hand, relativity is a {\em deterministic\/} theory equipped with a {\em dynamic\/} spacetime; on the other hand, quantum theory is a {\em probabilistic\/} theory embedded in a {\em fixed\/} spacetime.
This suggests that quantum gravity is relaxed in both aspects, {\it i.e.,}~it is a {\em probabilistic\/} theory equipped with a {\em dynamic\/} spacetime~\cite{Hardy:2007bk}.
Quantum non-local correlations~\mbox{\cite{Einstein:1935rr,Bell:1964ws,Brunner:2014kr}} motivate this study since the possibility of a satisfactory causal explanation~\cite{Reichenbach:1956vl} for such correlations is questionable~\cite{Bell:1964ws,Suarez:1997ds,Stefanov:2002da,Scarani:2002br,Coretti:2011ks,Bancal:2012jb,Barnea:2013ev,Scarani:2014fe,Wood:2015jf}.
Dropping the notion of a global time or of an {\em a priori\/} spacetime --- as has been suggested from different fields of research~\mbox{\cite{Leibniz:2000tq,vw,Page:1983kx,Wootters:1984gt,Bombelli:1987ez,MauroDAriano:2011in,Erker:2014th,Vedral:2014vb,Giovannetti:2015vg,Rankovic:2015vx}} --- dissolves this paradox.
This can be achieved by defining causal relations based on free randomness (see Figure~\ref{fig:causality}) as opposed to defining free randomness based on causal relations~\mbox{\cite{Colbeck:2011hw,Ghirardi:2013bb}}.
Such an approach gives a dynamic character to causality; causal connections are not predefined but are derived from the observed correlations.

Relaxations of quantum theory where the assumption of a global time is dropped have recently been studied widely~\cite{Hardy:2005wj,Hardy:2007bk,Hardy:2009,Chiribella:2009bh,Chiribella:2012jg,Colnaghi:2012dv,Oreshkov:2012uh,Baumeler:2013wy,Chiribella:2013bk,Costa:2013vc,Baumeler:2014cw,Brukner:2014vo,Ibnouhsein:J6_08c31,Ibnouhsein:2014uz,Morimae:2014ik,Oreshkov:2014ui,Araujo:2015wd,Oreshkov:2015vs,araujo14b,Feix:2015ww,Oreshkov:2015ge} (see Ref.~\cite{Brukner:2014if} for a review).
Our work follows the spirit of an operational quantum framework for such correlations developed by Oreshkov, Costa, and Brukner~\cite{Oreshkov:2012uh}.
Some correlations appearing in their quantum framework --- for two parties or more --- cannot be simulated by assuming a predefined causal order of the parties.
Such correlations are termed {\em non-causal}.
Analogously to non-locality, non-causal correlations could be witnessed by violating so-called causal inequalities~\cite{Oreshkov:2012uh,Baumeler:2013wy,Baumeler:2014cw,araujo14b}.
All causal inequalities in the two-party scenario and for binary inputs and outputs are presented in Ref.~\cite{araujo14b}.
In a previous work~\cite{Baumeler:2014cw}, we showed that in the classical limit of the quantum framework, {\it i.e.}, if it is restricted to probability theory, {\em classical non-causal\/} correlations can arise as well.
This result holds for three parties or more.
In the present work we follow this path and give a {\em representation\/} of all classical --- as opposed to quantum --- processes without predefined causal order {\em as polytopes}.
Such a representation helps in optimizing winning strategies for causal games~\cite{Oreshkov:2012uh,araujo14b} --- the optimization problem can be stated as a linear program ---, and for finding new causal games.

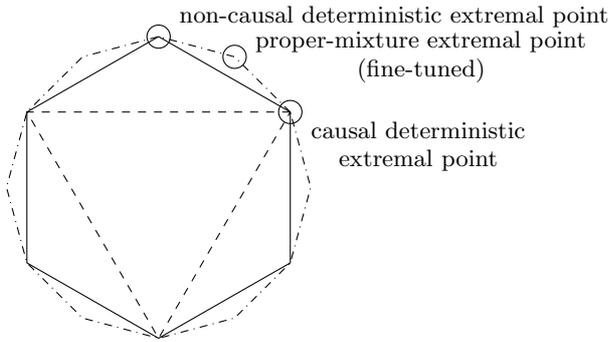
\begin{figure}
	\centering
	\begin{tikzpicture}[every text node part/.style={align=center}]
		\def\r{2}
		\draw[-,dashed] (30:\r) -- (150:\r) -- (270:\r) -- (30:\r);
		\draw[-] (30:\r) -- (90:\r) -- (150:\r) -- (210:\r) -- (270:\r) -- (330:\r) -- (30:\r);
		\draw[-,dashdotted] (30:\r) node[draw,solid,circle,label={-4:causal deterministic\\ extremal point}] (DC) {} -- (60:\r) node[draw,solid,circle,label={0:proper-mixture extremal point\\(fine-tuned)}] (PC) {} -- (90:\r) node[draw,solid,circle,label={3:non-causal deterministic extremal point}] (DNC) {} -- (120:\r) -- (150:\r) -- (180:\r) -- (210:\r) -- (240:\r) -- (270:\r) -- (300:\r) -- (330:\r) -- (0:\r) -- (30:\r);
	\end{tikzpicture}
	\caption{A qualitative representation of processes without predefined causal order studied in this work is given. The dashed region describes all processes that are achievable in a predefined causal order --- it also forms a polytope~\cite{Oreshkov:2015vs,araujo14b}. The polytope with the dashed-dotted lines is the polytope of processes without predefined causal order. The region in-between marked with the solid lines is the polytope of processes without predefined causal order restricted to deterministic extremal points.}
	\label{fig:polytopes}
\end{figure}

First, we present the framework of classical correlations without predefined causal order.
Then, we describe the polytope of processes that lead to such correlations implicitly and explicitly for scenarios with up to three parties and binary inputs and outputs.
In the general case, we give an implicit description of the polytope.
In addition, we construct the smaller polytope of classical processes without predefined causal order where all extremal points describe deterministic processes.
We call this polytope the {\em deterministic-extrema polytope}.
The processes from this polytope can be thought of as being ``more physical'' in the sense that its extremal points are not {\em proper mixtures\/} of logically inconsistent processes~\cite{Caslav}, {\it i.e.},~this set contains processes that can be written as a convex combination of deterministic ones from {\it within\/} the polytope only.
Our motivation for this is that some proper mixtures need to be fine-tuned~\cite{Wood:2015jf}, {\it i.e.}, tiny variations of the mixtures renders the processes logically inconsistent. 
The {\em fine-tuned\/} proper mixtures are the probabilistic extremal points of the larger polytope.
A qualitative representation of these polytopes is given in Figure~\ref{fig:polytopes}.

\section{Modelling classical correlations without predefined causal order}

\subsection{Causality, predefined causal order, and a framework of classical correlations without predefined causal order}
We describe an operational framework without global assumptions (other than logical consistency).
Causal relations are defined as in the interventionists' approach to causality~\cite{Price:1991kd,Woodward}: Outputs can be correlated to inputs and inputs are manipulated {\em freely\/} (see Figure~\ref{fig:causality}).
Defining causality based on {\em free randomness\/} is the converse approach to the one used in recent literature~\cite{Colbeck:2011hw,Ghirardi:2013bb}; there, free randomness is defined based on {\em causal relations}.
\begin{definition}[Causality~\cite{Baumeler:2014cw}]
	\rm
	For two correlated random variables~$X$ and~$A$, where~$X$ is an output and~$A$ is an input, {\it i.e.},~$A$ is chosen {\em freely}, we say that~$X$ is in the {\em causal future\/} of~$A$, or equivalently, that~$A$ is in the {\em causal past\/} of~$X$, denoted by~$X\succeq A$ or~$A\preceq X$.
	The negations of these relations are denoted by~$\not\succeq$ and~$\not\preceq$.
\end{definition}

Consider~$N$ parties~$\{S_j\}_{0\leq j< N}$, where party~$S_j$ has access to an input random variable~$A_j$ and generates an output random variable~$X_j$.
This allows us to causally order parties: If~$A_j$ is correlated to~$X_k$, then~$S_j$ is in the causal past of~$S_k$~($S_j\preceq S_k$).
To simplify the presentation, we write~$\vec X=(X_0,\dots,X_{N-1})$ and likewise for~$\vec A$,~$\vec O$, and~$\vec I$.

\begin{definition}[Two-party predefined causal order]
	\rm
	A {\em two-party predefined causal order\/} is a causal ordering of party~$S$ with input~$A$, output~$X$, and party~$T$ with input~$B$, output~$Y$, such that the distribution~$P_{X,Y|A,B}$ can be written as a convex combination of one-way signaling distributions
	\begin{align}
		P_{X,Y|A,B}=pP_{X|A,B,Y}P_{Y|B}+(1-p)P_{X|A}P_{Y|A,B,X}
		\,,
	\end{align}
	for some~$0\leq p\leq 1$.
\end{definition}
A definition for multi-party predefined causal order is given in Ref.~\cite{Oreshkov:2015vs}.
Such a definition turns out to be more subtle since a party~$S_j$ in the causal past of some other parties~$\{S_\ell\}_L$ can in principle influence everything in her causal future; in particular,~$S_j$ can influence the causal order of the parties~$\{S_\ell\}_L$.
We just state a Lemma that follows from such a definition and that is sufficient to prove our claims.
\begin{lemma}[Necesarry condition for predefined causal order]
	\label{def:predef}
	A necessary condition for a predefined causal order is that
	the probability distribution~$P_{\vec X|\vec I}$ can be written as a convex combination
	\begin{align}
		P_{\vec X|\vec I}=\sum_k p_kP_k
		\,,
	\end{align}
	with~$\sum_k p_k=1$ and~$\forall k:p_k\geq 0$, 
	such that in every distribution~$P_k$ at least one party is {\em not\/} in the causal future of any other party, {\it i.e.},
	\begin{align}
		\forall k\exists i\forall j\not =i:\,S_i\not\succeq^{P_k}S_j
		\,,
	\end{align}
	where~$\not\succeq^{P_k}$ stands for the causal relation that is deduced from the distribution~$P_k$.
\end{lemma}

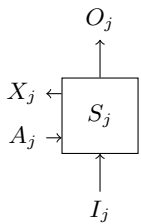
\begin{figure}
	\centering
	\begin{tikzpicture}
		\node[draw,rectangle,minimum width=1cm,minimum height=1cm] (S) {$S_j$};
		\draw[->] (S.150) -- ++(-0.2,0) node [left] {$X_j$};
		\draw[<-] (S.210) -- ++(-0.2,0) node [left] {$A_j$};
		\draw[->] (S.90) -- ++(0,0.5) node [above] {$O_j$};
		\draw[<-] (S.270) -- ++(0,-0.5) node [below] {$I_j$};
	\end{tikzpicture}
	\caption{A single party~$S_j$ describes a stochastic process~$P_{X_j,O_j|A_j,I_j}$.
The variables~$A_j$, and~$X_j$ model the input and the output.
The variable~$I_j$ is obtained from the environment~$E$; the party~$S_j$ feeds the variable~$O_j$ into the same environment.}
	\label{fig:party}
\end{figure}

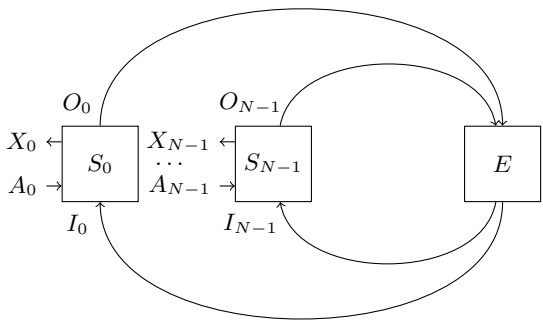
\begin{figure}
	\centering
        \begin{tikzpicture}
                \node[draw,rectangle,minimum height=1cm,minimum width=1cm] (S) {$S_0$};
                \draw[->] (S.150) -- ++(-0.2,0) node [left] {$X_0$};
                \draw[<-] (S.210) -- ++(-0.2,0) node [left] {$A_0$};
                \node[right=0.1cm of S] (D) {\dots};
                \node[draw,rectangle,minimum height=1cm,minimum width=1cm,right=0.5cm of D] (T) {$S_{N-1}$};
                \draw[->] (T.150) -- ++(-0.2,0) node [left] {$X_{N-1}$};
                \draw[<-] (T.210) -- ++(-0.2,0) node [left] {$A_{N-1}$};
                \node[draw,rectangle,minimum width=1cm,minimum height=1cm,right=2cm of T] (W) {$E$};
                \draw[->] (S) to[out=90,in=90] (W);
                \draw[<-] (S) to[out=270,in=270] (W);
                \draw[->] (T) to[out=80,in=100] (W);
                \draw[<-] (T) to[out=280,in=260] (W);
                \node (SO) at ([shift={(-0.3cm,0.8cm)}]S) {$O_0$};
                \node (SI) at ([shift={(-0.3cm,-0.8cm)}]S) {$I_0$};
                \node (TO) at ([shift={(-0.3cm,0.8cm)}]T) {$O_{N-1}$};
                \node (TI) at ([shift={(-0.3cm,-0.8cm)}]T) {$I_{N-1}$};
        \end{tikzpicture}
	\caption{The box~$E$ describes the environment. Because no predefined causal order is assumed between the parties, the random variable obtained by the parties is the result of~$E$ applied to the outgoing random variable of all parties. This picture combines states and channels, {\it i.e.}, signaling and no-signaling correlations. For example, assume that~$S_0$ is in the causal past all other parties.
	In that case, the random variable~$I_0$ is constant, whereas the random variable~$I_{j(\not=0)}$ could depend on~$A_0$.
For three parties or more, this framework gives rise to a new quality:~$E$ can describe a map where no~$I_j$ is a constant, yet where no contradiction arises. Such correlations are called non-causal. Similarly to the parties, the box~$E$ is a stochastic process~$P_{I_0,\dots,I_{N-1}|O_0,\dots,O_{N-1}}$.}
	\label{fig:nparties}
\end{figure}

In the framework without predefined causal order, each party~$S_j$ receives a random variable~$I_j$ from the environment~$E$ on which~$S_j$ can act.
After the interaction with~$I_j$, party~$S_j$ outputs a random variable~$O_j$ to the environment.
Both random variables~$I_j$ and~$O_j$ are {\em output\/} random variables.
The only {\em input\/} random variable a party has is~$A_j$.
The operation of~$S_j$ is a stochastic process mapping~$A_j,I_j$ to~$X_j,O_j$ (see~Figure~\ref{fig:party}).
A stochastic process is a probability distribution over the range conditioned on the domain; in this case, the stochastic process of party~$S_j$ (which in the following will also be called the {\em local operation\/} of party~$S_j$) is~$P_{X_j,O_j|A_j,I_j}$.
All parties are allowed to apply {\em any\/} possible operation described by probability theory.
Furthermore, they are isolated from each other, which means that they can interact only through the environment.
Because we do not make global assumptions (beyond logical consistency), the most general picture is that the random variables that are sent from the environment~$E$ to the parties are the result of a map on the random variables fed back by all parties to the same environment~$E$ (see Figure~\ref{fig:nparties}).
Such a composition of parties with the environment combines states and communication channels in one framework.

A party~$S_j$ has access to the four random variables~$X_j$,~$O_j$,~$I_j$, and~$A_j$, where~$A_j$ is chosen freely.
If we consider all parties together, we should get a probability distribution~$P_{\vec X,\vec I,\vec O|\vec A}$.
Furthermore, we ask the environment~$E$ to be a multi-linear functional of all local operations.
The motivation for this is that linear combinations of local operations should carry through to the probabilities~$P_{\vec X,\vec O,\vec I|\vec A}$.
This brings us to a definition of logical consistency.
\begin{definition}[Logical consistency]
	\label{def:lc}
	\rm
	An environment~$E$ is called {\em logically consistent\/} if and only if it is a multi-linear positive map on
	{\em any\/} choice of local operations~$\{P_{X_j,O_j|A_j,I_j}\}_{0\leq j<N}$ of all parties
	such that the composition of~$E$ with the local operations 
	results in a probability distribution~$P_{\vec X,\vec I,\vec O|\vec A}$.
\end{definition}
The linearity and positivity conditions from Definition~\ref{def:lc} imply Theorem~\ref{thm:env}, which states that the environment must be a stochastic process (conditional probability distribution).
\begin{theorem}[Logical consistent environment as stochastic process]
	\label{thm:env}
	The environment~$E$ is a stochastic process~$P_{\vec I|\vec O}$ that maps~$\vec O$ to~$\vec I$.
\end{theorem}
{\proof
	The environment is a multi-linear positive map~$\mathcal E$ on the probabilities (we omit the arguments for the sake of presentation)
	\begin{align}
		p_j:=P_{X_j,O_j|A_j,I_j}(x_j,o_j,a_j,i_j)
	\end{align}
	that party~$S_j$ outputs~$o_j$ to the environment and generates~$x_j$ conditioned on the setting~$a_j$ and on~$I_j=i_j$.
	Therefore, we write
	\begin{align}
		P_{\vec X,\vec I,\vec O|\vec A}(\vec x,\vec i,\vec o,\vec a)=\mathcal{E}(p_0,\dots,p_{N-1})
		\,.
	\end{align}
	Since~$\mathcal{E}$ is a multi-linear positive map and since it depends on~$\vec O$ and~$\vec I$ only, the above probability can be written as
	\begin{align}
		P_{\vec X,\vec I,\vec O|\vec A}(\vec x,\vec i,\vec o,\vec a)=E(\vec o,\vec i)p_0\cdots p_{N-1}
		\,,
		\label{eq:linearmap}
	\end{align}
	where~$E(\vec o,\vec i)$ is a number.
	This number must be non-negative, as otherwise the above expression~\eqref{eq:linearmap} is not a probability.
	By fixing~$\vec A=\vec a$ and by summing over~$\vec x$, we get
	\begin{align}
		P_{\vec I,\vec O|\vec A=\vec a}(\vec i,\vec o)&=\sum_{\vec x}E(\vec o,\vec i)p_0\cdots p_{N-1}\\
		&=E(\vec o,\vec i)\sum_{\vec x}p_0\cdots p_{N-1}\\
		&=E(\vec o,\vec i)p'_0\cdots p'_{N-1}
		\,,
	\end{align}
	where
	\begin{align}
		p'_j:=P_{O_j|I_j,A_j=a_j}(o_j,i_j)
		\,.
	\end{align}
	Let us fix the local operations~$p'_j$ of all parties to be
	\begin{align}
		P_{O_j|I_j,A_j=a_j}(o_j,i_j)=\begin{cases}
			1&o_j=0\,,\\
			0&\text{otherwise.}
		\end{cases}
	\end{align}
	From the total-probability condition we obtain
	\begin{align}
		\sum_{\vec o,\vec i}E(\vec o,\vec i)p'_0\cdots p'_{N-1}=\sum_{\vec i}E(\vec 0,\vec i)=1
		\,.
	\end{align}
	By repeating this calculation for different choices of local operations where the parties deterministically output a value, we get
	\begin{align}
		\forall \vec o:\,\sum_{\vec i}E(\vec o,\vec i)=1
		\,.
	\end{align}
	Therefore,~$E$ is a stochastic process~$P_{\vec I|\vec O}$.
\qed}

The following Corollary follows from Theorem~\ref{thm:env}.
\begin{corollary}
	A logical consistent environment~$P_{\vec I|\vec O}$ fulfills the property that
	under {\em any\/} choice of the local operations~$\{P_{X_j,O_j|A_j,I_j}\}_{0\leq j<N}$ of all parties, the expression~$P_{\vec I|\vec O}\prod_{j=0}^{N-1}P_{X_j,O_j|A_j,I_j}$ form a conditional probability distribution~$P_{\vec X,\vec I,\vec O|\vec A}$.
\end{corollary}

Note that not every conditional distribution~$P_{\vec I|\vec O}$ is logically consistent.
Some stochastic processes lead to {\em grandfather-paradox-type\/}~\cite{VoyGrandfather} inconsistencies.
Consider the following two extreme examples of such inconsistencies.
We describe the examples in the single-party scenario as depicted in Figure~\ref{fig:1party} and where~$O$,~$I$,~$X$, and~$A$ are binary random variables.
\begin{example}
	\rm
	Let the environment as well as the party~$S$ forward the random variable, {\it i.e.}, the operation of the environment is
	\begin{align}
		P_{I|O}(i,o)=\begin{cases}
			1&i=o\,,\\
			0&\text{otherwise,}
		\end{cases}
	\end{align}
	and the operation of the party~$S$ is
	\begin{align}
		P_{X,O|A,I}(x,o,a,i)=\begin{cases}
			1&o=i=x\,,\\
			0&\text{otherwise.}
		\end{cases}
	\end{align}
	Since the environment~$E$ and the party~$S$ forward the random variable, we have~$\Pr(O=I)=1$.
	However, it is unclear what value the probability~$P_O(0)$ should take.
	This is also known as the {\em causal-loop paradox}.
\end{example}
\begin{example}
	\rm
	We alter the local operation of party~$S$ to {\em negate\/} the binary random variable
	\begin{align}
		P_{X,O|A,I}(x,o,a,i)=\begin{cases}
			1&o=i\oplus 1=x\,,\\
			0&\text{otherwise.}
		\end{cases}
	\end{align}
	Now, we are faced with the {\em grandfather paradox}: if party~$S$ receives~$i=1$ from the environment, then she sends the value~$o=0$ to the environment. But in that case, she should receive~$i=0$ and not~$i=1$.
\end{example}

\subsection{Mathematical model of states, operations, evolution, and composition}
Let~$\{q_0,q_1,\dots\}$ be the sample space of a random variable~$Q$ with the probability measure~$P_Q$.
\begin{definition}[States, operations, evolution, and composition]
	\rm
	We represent a {\em state\/} corresponding to a random variable~$P_Q$ as the probability vector
\begin{align}
	\vec P_Q = (P_Q(q_0),P_Q(q_1),\dots)^T
	\,.
\end{align}
	A stochastic process~$P_{R|Q}$ from~$Q$ to a random variable~$R$ with sample space~$\{r_0,r_1,\dots\}$ describes an {\em operation\/} and is modeled by the stochastic matrix
\begin{align}
	\hat P_{R|Q}=
	\begin{pmatrix}
		P_{R|Q}(r_0,q_0) & P_{R|Q}(r_0,q_1) & \dots\\
		P_{R|Q}(r_1,q_0) & P_{R|Q}(r_1,q_1) & \dots\\
		\vdots  & \vdots  & \ddots
	\end{pmatrix}
	\,.
\end{align}
	The result~$P_R$ of {\em evolving\/} the random variable~$P_Q$ through the operation~$P_{R|Q}$ is given by the matrix multiplication
\begin{align}
	\vec P_R = \hat P_{R|Q} \vec P_Q
	\,.
\end{align}
	Finally, vectors and matrices are {\em composed\/} in parallel using the Kronecker product~$\otimes$.
\end{definition}
For example, by this definition, the output of a stochastic process~$P_{R|Q_0,Q_1}$ taking two inputs and producing one output is expressed by
\begin{align}
	\hat P_{R|Q_0,Q_1} \left( \vec P_{Q_0} \otimes \vec P_{Q_1} \right)
	\,.
\end{align}

\subsection{Set of logically consistent processes without predefined causal order}
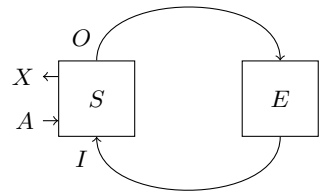
\begin{figure}
	\centering
	\begin{tikzpicture}
                \node[draw,rectangle,minimum height=1cm,minimum width=1cm] (S) {$S$};
                \draw[->] (S.150) -- ++(-0.2,0) node [left] {$X$};
                \draw[<-] (S.210) -- ++(-0.2,0) node [left] {$A$};
                \node[draw,rectangle,minimum width=1cm,minimum height=1cm,right=1.4cm of S] (W) {$E$};
                \draw[->] (S) to[out=90,in=90] (W);
                \draw[<-] (S) to[out=270,in=270] (W);
                \node (SO) at ([shift={(-0.2cm,0.8cm)}]S) {$O$};
                \node (SI) at ([shift={(-0.2cm,-0.8cm)}]S) {$I$};
        \end{tikzpicture}
	\caption{Party~$S$ is described by~$P_{X,O|A,I}$ and the environment~$E$ is~$P_{I|O}$.}
	\label{fig:1party}
\end{figure}

We derive the conditions on the environment~$E$ (stochastic process) such that it is {\em logically consistent}.
For simplicity, we start with the single-party scenario as depicted in Figure~\ref{fig:1party}; the party is denoted by~$S$ and the environment by~$E$.
We can further simplify our picture by fixing the value of~$A$ to~$a$ and by summing over~$X$:
\begin{align}
	\sum_x P_{X=x,O|A=a,I} = P_{O|I}
	\,.
\end{align}
The stochastic process of the environment~$E$ is~$P_{I|O}$.
For now, let us assume that~$S$ performs a deterministic operation~$D_{O|I}$.
This assumption is dropped later.
The operation applied by~$S$ can be written as a function
\begin{align}
	o = f(i)
	\,,
\end{align}
where~$i$ is a deterministic input value.
By embedding~$f$ into the process of~$E$, we get
\begin{align}
	P_{I|O}\left(i,f\left(i\right)\right)
	\,.
\end{align}
This can be interpreted as a probability measure of party~$S$ receiving the value~$i$ from the environment~$E$:
\begin{align}
	Q\left(i\right) = P_{I|O}\left(i,f\left(i\right)\right)
	\,.
\end{align}
For~$Q(i)$ to represent a probability measure, the values of~$Q$ for every deterministic value~$i$ must be non-negative and have to sum up to 1:
\begin{align}
	\forall i:\,Q\left( i \right) & \geq 0\,,\\
	\sum_{i} Q\left(i\right) &= 1
	\,.
\end{align}
We express both conditions in the matrix picture.
Non-negativity is achieved whenever all entries of the matrix~$\hat P_{I|O}$ are non-negative.
The total-probability conditions are formulated in the following way.
The value~$f(i)$ that is fed into the environment~$E$ is
\begin{align}
	\hat D_{O|I=\vec i} = \hat D_{O|I} \vec i
	\,.
\end{align}
The matrix~$\hat P_{I|O}$ fixed to providing the state~$\vec i$ to the party~$S$ is
\begin{align}
	\hat P_{I=\vec i|O} = \vec i^{\,T}\hat P_{I|O}
	\,.
\end{align}
Therefore, the probability of party~$S$ observing~$i$ is
\begin{align}
	Q\left(i\right) = \vec i^{\,T}\hat P_{I|O}\hat D_{O|I}\vec i
	\,,
\end{align}
and the law of total probability requires
\begin{align}
	\Tr\left(\hat P_{I|O}\hat D_{O|I}\right) = 1
	\,.
\end{align}
This condition remains the same if we relax the input to a stochastic input and the operation of~$S$ to a stochastic process~$P_{O|I}$.
The reason for this is that any stochastic input can be written as a convex combination of deterministic inputs,
and any stochastic process can be written as a convex combination of deterministic operations.
Therefore, the logical-consistency requirement asks the environment~$E$ to be restricted to those processes~$\hat E$ where, under {\em any choice\/} of the local operation~$P_{O|I}$ of party~$S$, the law of total probability
\begin{align}
	\Tr\left(\hat E \hat P_{O|I}\right) = 1\label{eq:condition}
\end{align}
and the non-negativity condition
\begin{align}
	\forall i,j:\, \hat E_{i,j}\geq 0\label{eq:condition2}
\end{align}
hold.
Because a stochastic process can be written as a convex mixture of deterministic operations, it is sufficient to ask for
\begin{align}
	\forall\hat D\in\mathcal{D}:\,&\Tr\left(\hat E \hat D\right) = 1\,\\
	\forall i,j:\,&\hat E_{i,j}\geq 0
\end{align}
for every operation~$\hat D$ from the set~$\mathcal{D}$ of all deterministic operations.
Thanks to linearity, we can straightforwardly extend these requirements to multiple parties, and arrive at Theorems~\ref{thm:totprobab} and~\ref{thm:nonnegprobab}.
\begin{theorem}[Total probability]
	\label{thm:totprobab}
	The law that the sum of the probabilities over the exclusive states the parties receive is~1 is satisfied if and only if
\begin{align}
	\forall\hat D_0,\hat D_1,\dots\in\mathcal{D}:\,\Tr(\hat E(\hat D_0 \otimes \hat D_1 \otimes \cdots)) = 1
	\,,
\end{align}
where~$\hat D_j$ represents a deterministic operation of party~$S_j$.
\end{theorem}

\begin{theorem}[Non-negative probabilities]
	\label{thm:nonnegprobab}
	The law that the probability of the parties observing a state is non-negative is satisfied if and only if
\begin{align}
	\forall i,j:\,\hat E_{i,j}\geq 0
	\,.
\end{align}
\end{theorem}

\subsection{Equivalence to the quantum correlations framework in the classical limit}
The ingredients of the framework by Oreshkov, Costa, and Brukner~\cite{Oreshkov:2012uh} are process matrices and local operations --- described by matrices as well.
All the matrices are completely-positive trace-preserving quantum maps in the Choi-Jamio{\l}kowski~\cite{Jamioikowski:1972ke,Choi:1975el} picture.
In the classical limit, the matrices become diagonal in the computational basis~\cite{Oreshkov:2012uh,Baumeler:2014cw}.
In the single-party scenario, the process matrix~$W$ is a map from the Hilbert space~$\mathcal{H}_O$ to the Hilbert space~$\mathcal{H}_I$.
The party's local operation~$A$ then again is a map from the Hilbert space~$\mathcal{H}_I$ to the Hilbert space~$\mathcal{H}_O$.
The conditions a process matrix~$W$ in a single-party scenario has to fulfill~\cite{Oreshkov:2012uh} are
\begin{align}
	\forall A\in\mathcal{M}:\,\Tr(WA)&=1\,,\label{eq:condocb1}\\
	W &\ge 0\,,\label{eq:condocb2}
\end{align}
where~$\mathcal{M}$ is the set of all completely-positive trace-preserving maps from the space~$\mathcal{H}_I$ to the space~$\mathcal{H}_O$.
Intuitively, the condition given by Equation~\eqref{eq:condocb1} ``short-circuits'' both maps and enforces the probabilities of the outcomes to sum~up~to~$1$.
\begin{theorem}[Equivalence]
	The quantum framework given by Equations~\eqref{eq:condocb1} and~\eqref{eq:condocb2} in the classical limit is equivalent to the description of classical correlations without predefined causal order given by Equations~\eqref{eq:condition} and~\eqref{eq:condition2}.
\end{theorem}
{\proof
The {\em process matrix\/}~$W$ in the quantum framework corresponds to the {\em stochastic process of the environment\/}~$E$ in our framework, and the {\em local operations\/} correspond to the {\em stochastic process of the parties}.
We show a bijection between process matrices and stochastic processes of the environment, and between local operations and stochastic processes of the parties.

A stochastic matrix~$\hat E$, representing the environment~$E$ in our framework, can be translated into the quantum framework by
\begin{align}
	W_{\hat E} = \sum_k \ket k\bra k_{\mathcal{H}_O} \otimes d\left(\hat E\ket k\sum_\ell\bra \ell_{\mathcal{H}_I}\right)
	\,,
\end{align}
where~$\ket k$ and~$\ket\ell$ are computational-basis states of the same dimension as~$\hat E$, and where the subscripts denote the respective Hilbert spaces.
This completely-positive trace-preserving map (expressed in the Choi-Jamio{\l}kowski picture) acts in the same way as the stochastic matrix~$\hat E$\/: The state~$\ket k$ is mapped to~$\hat E\ket k$.
The function~$d(\rho)$ takes the matrix~$\rho$ and cancels all off-diagonal terms, {\it i.e.},
\begin{align}
	d(\rho) = \sum_m \ket m\bra m \rho\ket m\bra m
	\,.
\end{align}
We can rewrite~$W_{\hat E}$ as
\begin{align}
	W_{\hat E} =\sum_k \ket k\bra k_{\mathcal{H}_O} \otimes \sum_m \ket m\bra m \hat E \ket k \bra m_{\mathcal{H}_I}
	\,.
\end{align}
Analogously, the stochastic matrix~$\hat P_{O|I}$ of the party can be translated into the quantum framework and becomes
\begin{align}
	A_{\hat P_{O|I}} = \sum_{k',m'} \ket {m'}\bra {m'} \hat P_{O|I}\ket {k'}\bra {m'}_{\mathcal{H}_O} \otimes \ket {k'}\bra {k'}_{\mathcal{H}_I}
	\,.
\end{align}
The reverse direction of the bijection follows from the description above.

Now, we show that the conditions~\eqref{eq:condocb1} and~\eqref{eq:condocb2} in a single-party scenario on a process matrix~$W$ co\"{\i}ncide with the conditions~\eqref{eq:condition} and~\eqref{eq:condition2} in our framework.
The non-negativity condition~\eqref{eq:condocb2} forces the probabilities of the outputs of~$W$ to be non-negative; the same holds for the condition~\eqref{eq:condition2} in our framework.
That the condition~\eqref{eq:condocb1} co\"{\i}ncides with the condition~\eqref{eq:condition} is shown below.
Forcing~$W$ and~$A$ to be diagonal in the computational basis gives
\begin{align}
	\Tr(WA) &= \sum_{i,j}\bra{i,j}WA\ket{i,j}\\
	&= \sum_{i,j}\bra{i,j}W\ket{i,j}\bra{i,j}A\ket{i,j}
	\,.
\end{align}
Substituting~$W$ with~$W_{\hat E}$ and~$A$ with~$A_{\hat P_{O|I}}$ yields
\begin{align}
	\sum_{i,j,m,k,m',k'}
	&\braket{i}{k}\braket{k}{i}
	\braket{j}{m}\bra{m} \hat E \ket{k}\braket{m}{j} \times\\
	\braket{i}{m'}&\bra{m'} \hat P_{O|I} \ket{k'}\braket{m'}{i}
	\braket{j}{k'}\braket{k'}{j}\\
	&=\sum_{i,j}
	\bra{j} \hat E \ket{i}\bra{i} \hat P_{O|I} \ket{j}\\
	&=\Tr\left(\hat E \hat P_{O|I}\right)
	\,,
\end{align}
which proves the claim.
The multi-party case follows through linearity.
\qed
}

\section{Polytope of classical processes without predefined causal order}
\subsection{Polytopes}
Convex polytopes can be represented in two different ways: The {\em $H$-representation\/} is a list of half-spaces where the intersection is the polytope, and the {\em $V$-representation\/} is a list of the extremal points of the polytope.
Algorithms like the double-description method~\cite{MRTT53,FP96} enumerate all extremal points of the polytope given the \mbox{$H$-representation}.
We used \textrm{cdd+}~\cite{cdd} for vertex enumeration.
The inverse problem is solved by its dual: a convex-hull algorithm.

\begin{figure}
	\centering
	\begin{tikzpicture}[scale=2]
		\draw[-] (0, -0.632456)
			-- (-0.316228, -0.316228)
			-- (-0.632456, 0)
			-- (-0.948683, 0.948683) node (f10a) {}
			-- (0.948683, 0.948683) node (f10b) {}
			-- (0.632456, 0)
			-- (0.316228, -0.316228)
			-- cycle;
		\draw[-,dashdotted] (f10a)
			-- (0,1.26491)
			-- (f10b);
		\node[fill,circle,black,inner sep=1pt,label=above:$C$] at (1.26491, 1.26491) {};
		\node[fill,circle,black,inner sep=1pt,label=above:$\bar C$] at (-1.26491, 1.26491) {};
		\node[fill,circle,black,inner sep=1pt,label=above:$\hat E_\text{ex1}$] at (0,1.26491) {};
		\node[fill,circle,black,inner sep=1pt,label=right:$\hat E_{\text{det}1}$] at (f10b) {};
	\end{tikzpicture}
	\caption{Here, we see a projection of the polytope of classical processes without predefined causal order among three parties and with binary inputs/outputs.
		The circular identity channel~$C$ and the circular bit-flip channel~$\bar C$ are logically inconsistent; they can be used to reproduce the grandfather's paradox.
	The solid lines mark the deterministic-extrema polytope and the dashed-dotted lines mark the additional space of logically consistent processes.
	Point~$\hat E_{\text{ex}1}$ is an extremal point of the polytope and is a uniform mixture of the deterministic processes~$C$ and~$\bar C$. The behavior of this point is shown in Figure~\ref{fig:exampleprobab}.
	Point~$\hat E_{\text{det}1}$ is an extremal point of the deterministic-extrema polytope, and is described in Figure~\ref{fig:examples}.
}
	\label{fig:slice}
\end{figure}
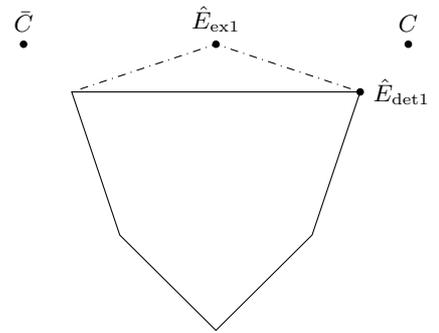

Here, we derive the polytope of classical processes without predefined causal order.
This polytope is represented by the dashed-dotted lines in Figure~\ref{fig:polytopes}.
A projection of the polytope for three parties and binary inputs/outputs onto a plane is given in Figure~\ref{fig:slice}.

\subsection{Single party, binary input, and binary output}
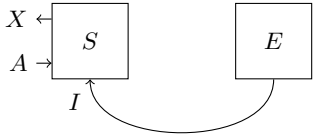
\begin{figure}
	\centering
	\begin{tikzpicture}
                \node[draw,rectangle,minimum height=1cm,minimum width=1cm] (S) {$S$};
                \draw[->] (S.150) -- ++(-0.2,0) node [left] {$X$};
                \draw[<-] (S.210) -- ++(-0.2,0) node [left] {$A$};
                \node[draw,rectangle,minimum width=1cm,minimum height=1cm,right=1.4cm of S] (W) {$E$};
                \draw[<-] (S) to[out=270,in=270] (W);
                \node (SI) at ([shift={(-0.2cm,-0.8cm)}]S) {$I$};
        \end{tikzpicture}
	\caption{All logically consistent single-party correlations that can be obtained with a feedback channel (see Figure~\ref{fig:1party}) can be simulated without feedback channel.}
	\label{fig:causal}
\end{figure}

We start with the polytope for one party (see Figure~\ref{fig:1party}) with a binary input and a binary output.
In this case, a process is described by a square matrix of dimension~$2$.
The most general process of the environment~$E$ is
\begin{align}
	\hat E=\hat P_{I|O} =
	\begin{pmatrix}
		w_0 & w_1\\
		w_2 & w_3\\
	\end{pmatrix}
	\,,
\end{align}
consisting of~$4$ variables.
The deterministic operations party~$S$ can apply are
\begin{align}
	\hat D_0=
	\begin{pmatrix}
		1 & 1\\
		0 & 0
	\end{pmatrix}
	\,,&\qquad
	\hat D_1=
	\begin{pmatrix}
		0 & 0\\
		1 & 1
	\end{pmatrix}
	\,,\\
	\hat D_2=
	\begin{pmatrix}
		1 & 0\\
		0 & 1
	\end{pmatrix}
	\,,&\qquad
	\hat D_3=
	\begin{pmatrix}
		0 & 1\\
		1 & 0
	\end{pmatrix}
	\,,
\end{align}
where~$\hat D_0$,~$\hat D_1$ produce a constant~$0$,~$1$, respectively, and where the matrix~$\hat D_2$ is the identity and~$\hat D_3$ the negation.
The equalities
\begin{align}
	\Tr\left( \hat E \hat D_0 \right) &= 1\,,\label{eq:cond1}\\
	\Tr\left( \hat E \hat D_1 \right) &= 1\,,\label{eq:cond2}\\
	\Tr\left( \hat E \hat D_2 \right) &= 1\,,\label{eq:cond3}
\end{align}
enforce
\begin{align}
	\Tr\left( \hat E \hat D_3 \right) &= 1
	\,.
\end{align}
This is shown as follows:
\begin{align}
	\Tr&\left( \hat E \hat D_0 \right)+ \Tr\left( \hat E \hat D_1 \right)+ \Tr\left( \hat E \hat D_2 \right)\\
	&=(w_0+w_2)+(w_1+w_3)+(w_0+w_3)\\
	&=2(w_0+w_3)+w_1+w_2\\
	&=2 \Tr\left( \hat E \hat D_2 \right) + \Tr\left( \hat E \hat D_3 \right)
	\,.
\end{align}
By eliminating three variables using the total-probability conditions~\eqref{eq:cond1},~\eqref{eq:cond2}, and~\eqref{eq:cond3} from above, we get
\begin{align}
	\hat P_{I|O} =
	\begin{pmatrix}
		w_0 & w_0\\
		1-w_0 & 1-w_0
	\end{pmatrix}
\end{align}
with the non-negativity conditions
\begin{align}
	w_0&\geq 0\,,\\
	1-w_0&\geq 0
	\,.
\end{align}
This solution set is a one-dimensional polytope with the extremal points~$0$ and~$1$.
All solutions describe a state.
This implies that all correlations that can be obtained in this framework with a single party and binary input and output, can also be obtained in a framework without feedback, {\it i.e.}, these correlations can be obtained causally~(see Figure~\ref{fig:causal}).

\subsection{Two parties, binary inputs, and binary outputs}
In the two-party case with a binary input and a binary output for each party, the process~$\hat E=\hat P_{I_0,I_1|O_0,O_1}$ of the environment is described by a square matrix of dimension~$2^2$.
The conditions are
\begin{align}
	\forall i,j\in\{0,1,2\}:\,
	\Tr\left( \hat E \left( \hat D_i\otimes  \hat D_j \right) \right) &= 1\,,\label{eq:cond2parties}\\
	\forall i,j:\, \hat E_{i,j} &\geq 0\label{eq:cond2parties2}
	\,.
\end{align}
With a similar argument as above, one can show that the operation~$\hat D_3$ does not need to be considered for either party.
The matrix~$\hat E$ consists of~$4^2$ unknowns, out of which~$3^2$ are eliminated by the total-probability conditions given by Equation~\eqref{eq:cond2parties}.
Thus, we are left with~$7$ unknowns, forming a~$7$-dimensional polytope with~$16$ inequalities.

The resulting $V$-representation of the polytope consists of~$12$ extremal points, all of which represent deterministic processes:
\begingroup
\allowdisplaybreaks
\begin{align}
	\hat E_0 = \begin{pmatrix}1&1&1&1\\0&0&0&0\\0&0&0&0\\0&0&0&0\end{pmatrix}\,,&\qquad
	\hat E_1 = \begin{pmatrix}0&0&0&0\\1&1&1&1\\0&0&0&0\\0&0&0&0\end{pmatrix}\,,\\
	\hat E_2 = \begin{pmatrix}0&0&0&0\\0&0&0&0\\1&1&1&1\\0&0&0&0\end{pmatrix}\,,&\qquad
	\hat E_3 = \begin{pmatrix}0&0&0&0\\0&0&0&0\\0&0&0&0\\1&1&1&1\end{pmatrix}\,,\\
	\hat E_4 = \begin{pmatrix}1&1&0&0\\0&0&1&1\\0&0&0&0\\0&0&0&0\end{pmatrix}\,,&\qquad
	\hat E_5 = \begin{pmatrix}0&0&1&1\\1&1&0&0\\0&0&0&0\\0&0&0&0\end{pmatrix}\,,\\
	\hat E_6 = \begin{pmatrix}0&0&0&0\\0&0&0&0\\1&1&0&0\\0&0&1&1\end{pmatrix}\,,&\qquad
	\hat E_7 = \begin{pmatrix}0&0&0&0\\0&0&0&0\\0&0&1&1\\1&1&0&0\end{pmatrix}\,,\\
	\hat E_8 = \begin{pmatrix}1&0&1&0\\0&0&0&0\\0&1&0&1\\0&0&0&0\end{pmatrix}\,,&\qquad
	\hat E_9 = \begin{pmatrix}0&1&0&1\\0&0&0&0\\1&0&1&0\\0&0&0&0\end{pmatrix}\,,\\
	\hat E_{10} = \begin{pmatrix}0&0&0&0\\1&0&1&0\\0&0&0&0\\0&1&0&1\end{pmatrix}\,,&\qquad
	\hat E_{11} = \begin{pmatrix}0&0&0&0\\0&1&0&1\\0&0&0&0\\1&0&1&0\end{pmatrix}
	\,.
\end{align}
\endgroup
In the following, we use~$A=S_0$ and~$B=S_1$.
The first four processes~$\hat E_0,\hat E_1,\hat E_2,\hat E_3$ represent the four constants~$(0,0),(0,1),(1,0),(1,1)$ as inputs to the parties~$A$ and~$B$. 
The next four processes represent a constant input to party~$A$ (processes~$\hat E_4$ and~$\hat E_5$ produce the constant~$0$, and the other two processes produce the constant~$1$) and a channel from party~$A$ to party~$B$; the processes~$\hat E_4$ and~$\hat E_6$ describe the identity channel, and~$\hat E_5$ and~$\hat E_7$ describe the bit-flip channel.
The last four processes are analogous, with a channel from~$B$ to~$A$ and where party~$B$ receives a constant.
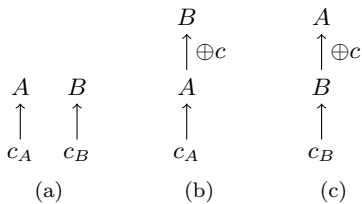
\begin{figure}
	\centering
	\subfloat[\label{fig:2pconst}]{
		\begin{tikzpicture}
			\node (A) {$A$};
			\node[right=0.5cm of A.center] (B) {$B$};
			\draw[<-] (A) -- ++(0,-0.7) node[below] (VA) {$c_A$};
			\draw[<-] (B) -- ++(0,-0.7) node[below] (VB) {$c_B$};
		\end{tikzpicture}
	}
	\qquad
	\subfloat[\label{fig:2pafirst}]{
		\begin{tikzpicture}
			\node (A) {$A$};
			\node[above=0.7cm of A.center] (B) {$B$};
			\draw[<-] (A) -- ++(0,-0.7) node[below] (VA) {$c_A$};
			\draw[->] (A) -- (B) node[midway,right] (N) {$\oplus c$};
		\end{tikzpicture}
	}
	\qquad
	\subfloat[\label{fig:2pbfirst}]{
		\begin{tikzpicture}
			\node (A) {$B$};
			\node[above=0.7cm of A.center] (B) {$A$};
			\draw[<-] (A) -- ++(0,-0.7) node[below] (VA) {$c_B$};
			\draw[->] (A) -- (B) node[midway,right] (N) {$\oplus c$};
		\end{tikzpicture}
	}
	\caption{(a)~Both parties~$A$ and~$B$ receive a constant each. (b)~Party~$A$ receives a constant and sends a bit through the identity~$(c=0)$ or the bit-flip~$(c=1)$ channel to~$B$. (c)~Same as~(b), where the parties are interchanged.}
	\label{fig:16ch}
\end{figure}
All these~$12$ processes act deterministically on bits for two parties where at least one party receives a constant (see Figure~\ref{fig:16ch}).
Therefore, every such channel can be simulated in a causal fashion.
This result generalized to higher dimensions was already shown by taking the classical limit of the framework for quantum correlations without predefined causal order~\cite{Oreshkov:2012uh}.

\subsection{Three parties, binary inputs, and binary outputs}
\label{sec:3pbin}
The process of the environment~$E$ in a three party setup with binary inputs and outputs is described by a square matrix~$\hat E=\hat P_{I_0I_1I_2|O_0O_1O_2}$ of dimension~$2^3$.
The matrix~$\hat E$ consists of~$4^3$ variables, out of which~$3^3$ can be eliminated with the total-probability conditions
\begin{align}
	\forall i,j,k\in\{0,1,2\}: \Tr\left( \hat E \left( \hat D_i\otimes  \hat D_j \otimes \hat D_k\right) \right) = 1
	\label{eq:3ptp}
	\,,
\end{align}
resulting in a~$37$-dimensional polytope with~$4^3$ linear constraints (non-negative probabilities):
\begin{align}
	\forall i,j:\,
	\hat E_{i,j} \geq 0
	\,.
	\label{eq:3pnn}
\end{align}
Solving this polytope yields~$710'760$ extremal points.
Only~$744$ extremal points out of these~$710'760$ are deterministic, {\it i.e.}, consist of 0-1 values; the remaining extremal points are so-called {\em proper mixtures\/} of logically inconsistent processes.
Such proper mixtures are {\em not\/} convex combinations of deterministic extremal points {\em inside\/} the polytope, but are convex combinations of deterministic points where some lie {\em outside\/} of the polytope --- any process from outside of the polytope leads to logical inconsistencies.
Interestingly, this smaller polytope (hence, also the polytope described by the Equations~\eqref{eq:3ptp} and~\eqref{eq:3pnn}) consists of processes that cannot be simulated using a predefined causal order, {\it i.e.}, processes where {\em no party receives a constant}, implying that {\em every party causally succeeds some other party}.
The~$744$ deterministic extremal points are discussed in Section~\ref{sec:detpoly} along with the general polytope restricted to the deterministic extremal points.

\subsection{General case}
We describe the polytope for logically consistent classical processes without predefined causal order in the general case.
Let~$n$ be the number of parties and let~$d$ be the dimension of the states entering and leaving every laboratory. 
This leaves us with a~$d^n\times d^n$ stochastic matrix~$\hat E$ describing the environment.
Every party can perform an operation that is a convex mixture of all~$d^d$ deterministic operations.
The set of all deterministic operations is denoted by~$\mathcal{D}$.
For every party, under any choice of deterministic operation~$D\in\mathcal{D}$, the trace of the environment~$\hat E$ multiplied with the local operations is constrained to give~$1$ (see Theorem~\ref{thm:totprobab}).
However --- as in the binary-input/output case above \mbox{---,} some of these constraints are redundant.
\begin{theorem}[Sufficient set for total-probability conditions]
	The total-probability conditions to this family of operations
	\begin{align}
		\hat D_{i,j}=
		\begin{pmatrix}
			1 & 1 & 1 & \dots & 1 & 0 & 1 & \dots & 1\\
			0 & 0 & 0 & \dots & 0 & 0 & 0 & \dots & 0\\
			\vdots & \vdots & \vdots & \ddots & \vdots & \vdots & \vdots & \ddots & \vdots\\
			0 & 0 & 0 & \dots & 0 & 0 & 0 & \dots & 0\\
			0 & 0 & 0 & \dots & 0 & 1 & 0 & \dots & 0\\
			0 & 0 & 0 & \dots & 0 & 0 & 0 & \dots & 0\\
			\vdots & \vdots & \vdots & \ddots & \vdots & \vdots & \vdots & \ddots & \vdots\\
			0 & 0 & 0 & \dots & 0 & 0 & 0 & \dots & 0
		\end{pmatrix}
		\,,
	\end{align}
	where~$j$ is output for input~$i$ and~$0$ otherwise,
	imply the total-probability conditions for all remaining deterministic operations of the same dimension, {\it i.e.},
	\begin{align}
		\forall &i_0,j_0,i_1,j_j,\dots,i_{n-1},j_{n-1}\geq0:\\
		&\Tr\left(\hat E\left( \hat D_{i_0,j_0}\otimes\dots\otimes \hat D_{i_{n-1},j_{n-1}} \right)\right)=1\notag\\
		\implies\\
		\forall &\hat D_0,\dots,\hat D_{n-1}\in\mathcal{D}:\\
		&\Tr\left(\hat E\left( \hat D_0\otimes\dots\otimes \hat D_{n-1} \right)\right)=1
		\,.
	\end{align}
\end{theorem}
{\proof
We restrict ourselves to the single-party scenario --- the multi-party case follows through linearity.
Let~$\vec v_i$ be the~\mbox{$d$-dimensional} vector with a~$1$-entry at position~$i$ and~$0$'s everywhere else.
We can write a~\mbox{$d$-dimensional} matrix~$\hat D_{i,j}$ as
\begin{align}
	\hat D_{i,j}=\left(\sum_{m\not=i} \vec v_0\vec v_m^T\right)+\vec v_j\vec v_i^T
	\,.
\end{align}
A general deterministic matrix~$D\in\mathcal{D}$ of the same dimension, where~$k$ is mapped to~$a_k$, is expressed as
\begin{align}
	D=\sum_{k}\vec v_{a_k}\vec v_k^T
	\,.
\end{align}
On the one hand, using the antecedent above, we have
\begin{align}
	\Tr\left(\hat E\sum_{k:a_k\not=0}\hat D_{k,a_k}\right)
	=\sum_{k:a_k\not=0}\Tr\left( \hat E\hat D_{k,a_k} \right)=\ell
	\label{eq:sumid}
\end{align}
with~$\ell=|\{k|a_k\not=0 \}|$.
On the other hand, we can rewrite~$\sum_{k:a_k\not=0}\hat D_{k,a_k}$ as
\begin{align}
	\sum_{k:a_k\not=0}\hat D_{k,a_k} &= \sum_{k:a_k\not=0}\left(\sum_{m\not=k}\vec v_0\vec v_m^T\right)+\vec v_{a_k}\vec v_k^T\\
	&=\sum_{k:a_k\not=0}\vec v_{a_k}\vec v_k^T + \sum_{k:a_k=0}\vec v_0\vec v_k^T \\
	&\quad- \sum_{k:a_k=0}\vec v_0\vec v_k^T + \sum_{k:a_k\not=0}\sum_{m\not=k}\vec v_0\vec v_m^T\\
	&=D+\sum_{k:a_k\not=0}\sum_{m\not=k}\vec v_0\vec v_m^T-\sum_{k:a_k=0}\vec v_0\vec v_k^T\\
	&=D+\ell\sum_{k:a_k=0}\vec v_0\vec v_k^T+(\ell-1)\sum_{k:a_k\not=0}\vec v_0\vec v_k^T\notag\\
	&\quad-\sum_{k:a_k=0}\vec v_0\vec v_k^T\\
	&=D+(\ell-1)\sum_k\vec v_0\vec v_k^T\\
	&=D+(\ell-1)\hat D_{0,0}
	\,.
\end{align}
Therefore, 
\begin{align}
	\Tr\left(\hat E\sum_{k:a_k\not=0}\hat D_{k,a_k}\right)
	&=\Tr(\hat ED)+(\ell-1)\Tr\left( \hat E\hat D_{0,0} \right)\\
	&=\Tr(\hat ED)+\ell-1
	\,,
\end{align}
which, with the identity~\eqref{eq:sumid}, implies
\begin{align}
	\Tr(\hat ED)=1
	\,.
\end{align}
\qed}

The family~$\{\hat D_{i,j}|i,j\in I\}$ of deterministic operations with the set \mbox{$I=\{0,\dots,n-1\}$} has size~\mbox{$d(d-1)+1$}.
\begin{theorem}[Polytope]
	The~$H$-representation of the polytope of logically consistent classical processes without predefined causal order~is
	\begin{align}
		\forall \hat D_0,\hat D_1,\dots,\hat D_{n-1}\in \{\hat D_{i,j}\}_{I\times I}:\\
		\Tr\left( \hat E \left( \hat D_0\otimes  \hat D_1 \otimes \dots\otimes \hat D_{n-1}\right) \right) &= 1
		\,,\\
		\forall i,j:\,\hat E_{i,j}&\geq 0
		\,.
	\end{align}
	The polytope has~$d^{2n}$ facets and dimension
	\begin{align}
		d^{2n}-(d(d-1)+1)^n
		\,,
	\end{align}
which is exponential in the number of parties.
\end{theorem}

\section{The deterministic-extrema polytope}
\label{sec:detpoly}
\begin{definition}[Deterministic-extrema polytope]
	\rm
	The {\em deterministic-extrema polytope\/} is defined as the polytope of logically consistent processes without predefined causal order {\em where all extremal points are deterministic processes\/} (see polytope with the solid lines in Figure~\ref{fig:polytopes}).
\end{definition}

	The deterministic-extrema polytope excludes {\em proper mixtures\/} of logically inconsistent processes.
	Such mixtures (consistent mixture of inconsistent points) are convex combinations of deterministic points where at least one deterministic point is outside of the polytope.
To find this polytope, one can first solve the extremal points of the general polytope and thereafter select the boolean solutions.
These boolean solutions form the~\mbox{$V$-representation} of the polytope in discussion.

\subsection{Three parties, binary inputs, and binary outputs}
We discuss the deterministic-extrema polytope in the setting of three parties and binary inputs and outputs.
To simplify the presentation, we use~$A=S_0$,~$B=S_1$, \mbox{$C=S_2$}, $O_A=O_0$,~$O_B=O_1$,~$O_C=O_2$,~$I_A=I_0$,~$I_B=I_1$, and~$I_C=I_2$.
As described in Section~\ref{sec:3pbin}, this polytope has~$744$ extremal points.
They can be characterized as follows.

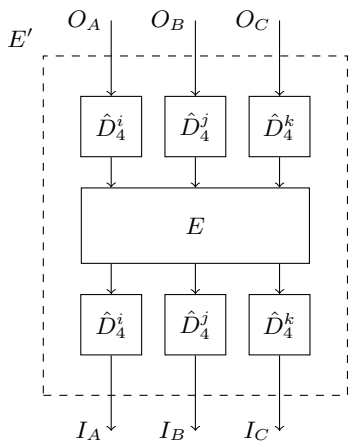
\begin{figure}
	\centering
	\begin{tikzpicture}
                \node[draw,rectangle,minimum width=3cm,minimum height=1cm] (W) {$E$};
		\node[draw,rectangle,minimum width=0.8cm,minimum height=0.8cm,above left=0.4cm and 0.7cm of W.north] (UL) {$\hat D_4^i$};
		\node[draw,rectangle,minimum width=0.8cm,minimum height=0.8cm,above=0.4cm of W.north] (UC) {$\hat D_4^j$};
		\node[draw,rectangle,minimum width=0.8cm,minimum height=0.8cm,above right=0.4cm and 0.7cm of W.north] (UR) {$\hat D_4^k$};
		\node[draw,rectangle,minimum width=0.8cm,minimum height=0.8cm,below left=0.4cm and 0.7cm of W.south] (BL) {$\hat D_4^i$};
		\node[draw,rectangle,minimum width=0.8cm,minimum height=0.8cm,below=0.4cm of W.south] (BC) {$\hat D_4^j$};
		\node[draw,rectangle,minimum width=0.8cm,minimum height=0.8cm,below right=0.4cm and 0.7cm of W.south] (BR) {$\hat D_4^k$};
		\draw[->] (UL.south) -- ++(0,-0.4cm);
		\draw[->] (UC.south) -- ++(0,-0.4cm);
		\draw[->] (UR.south) -- ++(0,-0.4cm);
		\draw[<-] (BL.north) -- ++(0,0.4cm);
		\draw[<-] (BC.north) -- ++(0,0.4cm);
		\draw[<-] (BR.north) -- ++(0,0.4cm);
		\draw[<-] (UL.north) -- ++(0,1cm) node[above,left] {$O_A$};
		\draw[<-] (UC.north) -- ++(0,1cm) node[above,left] {$O_B$};;
		\draw[<-] (UR.north) -- ++(0,1cm) node[above,left] {$O_C$};;
		\draw[->] (BL.south) -- ++(0,-1cm) node[below,left] {$I_A$};;
		\draw[->] (BC.south) -- ++(0,-1cm) node[below,left] {$I_B$};;
		\draw[->] (BR.south) -- ++(0,-1cm) node[below,left] {$I_C$};;
		\node[draw,dashed,rectangle,minimum width=4cm,minimum height=4.5cm,label=north west:$E'$] (Wp) {};
	\end{tikzpicture}
	\caption{By starting from a logically consistent environment~$E$ and for any choice of~$i,j,k\in\{0,1\}$, one can construct another logically consistent environment~$E'$.}
	\label{fig:embed}
\end{figure}

Assume that~$\hat E$, when the parties locally apply the identity operation, maps~$(0,0,0)$ to~$(0,0,0)$, {\it i.e.},~$(0,0,0)$ is a fixed-point.
Then, any other extremal point~$\hat E'$ is obtained by the local operations identity~$\hat D_3$ and bit-flip~$\hat D_4$, where we embed these local operations into the environment.
Let~$\hat L_{i,j,k}$ be the local operation of the three parties
\begin{align}
	\hat L_{i,j,k}=\hat D_4^i\otimes \hat D_4^j \otimes \hat D_4^k
	\,,
\end{align}
{\it i.e.}, party~$A$ performs the identity if~$i=0$ and the bit-flip operation if~$i=1$ --- the other parties' local operations are defined in the same way.
The extremal point~$\hat E'$ can be described as
\begin{align}
	\hat E'=\hat L_{i,j,k}\hat E\hat L_{i,j,k}
	\,,
\end{align}
where, as described above, the operations are embedded into the environment (see Figure~\ref{fig:embed}).
Logical consistency of the environment~$\hat E'$ follows because we started with a logically consistent~$\hat E$ and the operations act on single parties.
The process~$\hat E'$ maps~$(i,j,k$) to~$(i,j,k)$.
Thus, starting with~$\hat E$, for any choice of~$i,j,k$, we obtain a different extremal point.
There are~$2^3-1$ alternative extremal points that can be constructed in this fashion.
From this we conclude that~$744/8=93$ extremal points are such, that~$(0,0,0)$ is a fixed-point under locally applying the identity.
We restrict our analysis to these~$93$ extremal points; all others can be obtained by the above construction.
The following analysis is structured depending on the number of parties that receive a constant from the environment.

There exists only one extremal point where~$(0,0,0)$ is mapped to~$(0,0,0)$ under applying identity locally, and where {\em every party\/} receives a constant: the constant~$(0,0,0)$.

Assume exactly {\em two parties\/} receive a constant, which leaves us with three possibilities of choosing them.
Fix these parties to be~$A$ and~$B$.
The third party~$C$ receives a value that depends on the operation of~$A$ or of~$B$ or of both.
Thus, we are in the case~$A\preceq C$ or~$B\preceq C$.
Inevitably, the constant must be~$(0,0)$; otherwise the fixed-point~$(0,0,0)$ is not recovered.
Party~$C$ receives a value that depends on the value fed back by~$A$ and~$B$; there exist~$2^3-1=7$ such functions where we have excluded the constant and all operations where~$C$ receives a value different from~$0$ on inputs~$(0,0)$ to the environment from~$A$ and~$B$.
Therefore, under all permutations of the parties,~$21$ extremal points give a constant to two parties and have the fixed-point~$(0,0,0)$ when the identity is applied locally.

\begin{figure}
	\centering
	\subfloat[\label{fig:depconst}]{
		\begin{tikzpicture}
			\node (A) {$A$};
			\node[below=0.2cm of A.center] (B) {$B$};
			\node[below=0.2cm of B.center] (C) {$C$};
		\end{tikzpicture}
	}
	\quad
	\subfloat[\label{fig:dep2const}]{
		\begin{tikzpicture}
			\node (A) {$A$};
			\node[below right=0.1cm and 0.5cm of A.center] (C) {$C$};
			\node[below left=0.1cm and 0.5cm of C.center] (B) {$B$};
			\draw[->] (A) -- (C);
			\draw[->] (B) -- (C);
			\node[right=1.3cm of A.center] (A2) {$A$};
			\node[below right=0.1cm and 0.5cm of A2.center] (C2) {$B$};
			\node[below left=0.1cm and 0.5cm of C2.center] (B2) {$C$};
			\draw[->] (A2) -- (C2);
			\draw[->] (B2) -- (C2);
			\node[right=1.3cm of A2.center] (A3) {$B$};
			\node[below right=0.1cm and 0.5cm of A3.center] (C3) {$A$};
			\node[below left=0.1cm and 0.5cm of C3.center] (B3) {$C$};
			\draw[->] (A3) -- (C3);
			\draw[->] (B3) -- (C3);
		\end{tikzpicture}
	}//~//~%
	\subfloat[\label{fig:dep1const1}]{
		\begin{tikzpicture}
			\node (B) {$B$};
			\node[below left=0.1cm and 0.5cm of B.center] (A) {$A$};
			\node[below right=0.1cm and 0.5cm of A.center] (C) {$C$};
			\draw[->] (A) -- (B);
			\draw[->] (A) -- (C);
			\node[right=1.3cm of B.center] (B2) {$A$};
			\node[below left=0.1cm and 0.5cm of B2.center] (A2) {$B$};
			\node[below right=0.1cm and 0.5cm of A2.center] (C2) {$C$};
			\draw[->] (A2) -- (B2);
			\draw[->] (A2) -- (C2);
			\node[right=1.3cm of B2.center] (B3) {$A$};
			\node[below left=0.1cm and 0.5cm of B3.center] (A3) {$C$};
			\node[below right=0.1cm and 0.5cm of A3.center] (C3) {$B$};
			\draw[->] (A3) -- (B3);
			\draw[->] (A3) -- (C3);
		\end{tikzpicture}
	}

	\subfloat[\label{fig:dep1const2}]{
		\begin{tikzpicture}
			\node (A) {$A$};
			\node[right=0.5cm of A.center] (B) {$B$};
			\node[right=0.5cm of B.center] (C) {$C$};
			\draw[->] (A) -- (B);
			\draw[->] (B) -- (C);
			\node[below=0.3cm of A.center] (A2) {$A$};
			\node[right=0.5cm of A2.center] (B2) {$C$};
			\node[right=0.5cm of B2.center] (C2) {$B$};
			\draw[->] (A2) -- (B2);
			\draw[->] (B2) -- (C2);
			\node[right=1.3cm of C.center] (A3) {$B$};
			\node[right=0.5cm of A3.center] (B3) {$A$};
			\node[right=0.5cm of B3.center] (C3) {$C$};
			\draw[->] (A3) -- (B3);
			\draw[->] (B3) -- (C3);
			\node[below=0.3cm of A3.center] (A4) {$B$};
			\node[right=0.5cm of A4.center] (B4) {$C$};
			\node[right=0.5cm of B4.center] (C4) {$A$};
			\draw[->] (A4) -- (B4);
			\draw[->] (B4) -- (C4);
			\node[right=1.3cm of C3.center] (A5) {$C$};
			\node[right=0.5cm of A5.center] (B5) {$A$};
			\node[right=0.5cm of B5.center] (C5) {$B$};
			\draw[->] (A5) -- (B5);
			\draw[->] (B5) -- (C5);
			\node[below=0.3cm of A5.center] (A6) {$C$};
			\node[right=0.5cm of A6.center] (B6) {$B$};
			\node[right=0.5cm of B6.center] (C6) {$A$};
			\draw[->] (A6) -- (B6);
			\draw[->] (B6) -- (C6);
		\end{tikzpicture}
	}

	\subfloat[\label{fig:dep1const3}]{
		\begin{tikzpicture}
			\node (B) {$B$};
			\node[below left=0.1cm and 0.5cm of B.center] (A) {$A$};
			\node[below right=0.1cm and 0.5cm of A.center] (C) {$C$};
			\draw[->] (A) -- (B);
			\draw[->] (A) -- (C);
			\draw[->] (B) -- (C);
			\node[below=0.3cm of C.center] (B2) {$B$};
			\node[below left=0.1cm and 0.5cm of B2.center] (A2) {$A$};
			\node[below right=0.1cm and 0.5cm of A2.center] (C2) {$C$};
			\draw[->] (A2) -- (B2);
			\draw[->] (A2) -- (C2);
			\draw[->] (C2) -- (B2);
			\node[right=1.3cm of B.center] (B3) {$A$};
			\node[below left=0.1cm and 0.5cm of B3.center] (A3) {$B$};
			\node[below right=0.1cm and 0.5cm of A3.center] (C3) {$C$};
			\draw[->] (A3) -- (B3);
			\draw[->] (A3) -- (C3);
			\draw[->] (B3) -- (C3);
			\node[below=0.3cm of C3.center] (B4) {$A$};
			\node[below left=0.1cm and 0.5cm of B4.center] (A4) {$B$};
			\node[below right=0.1cm and 0.5cm of A4.center] (C4) {$C$};
			\draw[->] (A4) -- (B4);
			\draw[->] (A4) -- (C4);
			\draw[->] (C4) -- (B4);
			\node[right=1.3cm of B3.center] (B5) {$A$};
			\node[below left=0.1cm and 0.5cm of B5.center] (A5) {$C$};
			\node[below right=0.1cm and 0.5cm of A5.center] (C5) {$B$};
			\draw[->] (A5) -- (B5);
			\draw[->] (A5) -- (C5);
			\draw[->] (B5) -- (C5);
			\node[below=0.3cm of C5.center] (B6) {$A$};
			\node[below left=0.1cm and 0.5cm of B6.center] (A6) {$C$};
			\node[below right=0.1cm and 0.5cm of A6.center] (C6) {$B$};
			\draw[->] (A6) -- (B6);
			\draw[->] (A6) -- (C6);
			\draw[->] (C6) -- (B6);
		\end{tikzpicture}
	}

	\subfloat[\label{fig:dep1const4}]{
		\begin{tikzpicture}
			\node (B) {$B$};
			\node[below left=0.1cm and 0.5cm of B.center] (A) {$A$};
			\node[below right=0.1cm and 0.5cm of A.center] (C) {$C$};
			\draw[->] (A) -- (B);
			\draw[->] (A) -- (C);
			\draw[<->] (B) -- (C);
			\node[right=1.3cm of B.center] (B2) {$A$};
			\node[below left=0.1cm and 0.5cm of B2.center] (A2) {$B$};
			\node[below right=0.1cm and 0.5cm of A2.center] (C2) {$C$};
			\draw[->] (A2) -- (B2);
			\draw[->] (A2) -- (C2);
			\draw[<->] (B2) -- (C2);
			\node[right=1.3cm of B2.center] (B3) {$A$};
			\node[below left=0.1cm and 0.5cm of B3.center] (A3) {$C$};
			\node[below right=0.1cm and 0.5cm of A3.center] (C3) {$B$};
			\draw[->] (A3) -- (B3);
			\draw[->] (A3) -- (C3);
			\draw[<->] (B3) -- (C3);
		\end{tikzpicture}
	}

	\subfloat[\label{fig:3to3}]{
		\begin{tikzpicture}
			\node (B) {$B$};
			\node[below left=0.1cm and 0.5cm of B.center] (A) {$A$};
			\node[below right=0.1cm and 0.5cm of A.center] (C) {$C$};
			\draw[<->] (A) -- (B);
			\draw[<->] (A) -- (C);
			\draw[<->] (B) -- (C);
		\end{tikzpicture}
	}
	\caption{(a)~Every party receives a constant. For an environment with fixed point~$(0,0,0)$ when the parties locally apply the identity map, this constant must be~$(0,0,0)$.
		(b)~Two parties receive a constant~$(0,0)$, the third party receives a value depending on the other parties' state fed to the environment. For each of the three cases, seven different functions exist. (c)~For each of the three cases, the identity function only is consistent with the setup. (d)~Here as well, only the identity channel is consistent with the fixed-point. (e)~Five different functions are possible per setup. (f)~Here, eight functions per setup are consistent with the fixed-point. 
	(g)~No party receives a constant, yet no contradiction arises under any choice of local operations. For a fixed-point~$(0,0,0)$ where the parties locally apply the identity map, eight different functions that fulfill these requirements exist.
	In total, the number of deterministic extremal points where~$(0,0,0)$ is mapped to~$(0,0,0)$ when the parties apply the identity operation is $1\text{ (a)} + 3\cdot7\text{ (b)}+3\text{ (c)}+6\text{ (d)}+6\cdot5\text{ (e)}+3\cdot 8\text{ (f)}+8\text{ (g)}=93$. 
	}
	\label{fig:dep}
\end{figure}
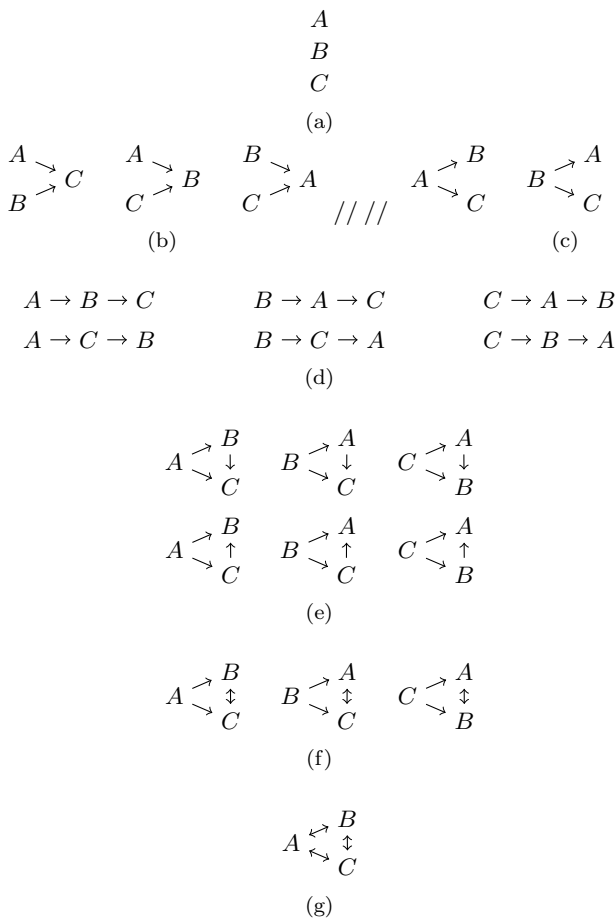

In a next step, assume that exactly {\em one party\/} receives a constant.
This assumption, again, allows for three different setups, as we can choose which party receives a constant.
Without loss of generality, let~$A$ be this party, {\it i.e.},~$A\preceq B$ and~$A\preceq C$; the constant must be~$0$ again in order to comply with the requirement of the fixed-point.
Now, we are left with several possibilities on how~$B$ and~$C$ depend on~$A$ and on each other.
As a first case, we assume that~$B$ and~$C$ do not depend on each other, but depend on~$A$ only ($A \preceq B$ and~$A\preceq C$).
This dependency cannot be different from the identity channels from~$A$ to~$B$ and from~$A$ to~$C$; the alternative would be the bit-flip channel that would not reproduce the desired fixed-point~$(0,0,0)$.
This gives us~$3$ different extremal points under all permutations of the parties.
Another possibility on the dependencies is that~$B$ depends on~$A$, and~$C$ depends on~$B$, {\it i.e.},~$A\preceq B\preceq C$, and the interchange of parties~$B$ and~$C$.
The channels --- by following the same reasoning above --- again must be the identity channels: This gives us~$6$ extremal points.
Now, we look at the case where~$B$ depends on~$A$ and where~$C$ depends on both,~$A$ and~$B$, {\it i.e.},~$A\preceq B$,~$A\preceq C$, and~$B\preceq C$, and any permutation of the parties.
There are~$6$ permutations.
The constant that~$A$ receives must be~$0$, party~$B$ must depend trivially on~$A$ (the identity channel) and party~$C$ can depend in~$5$ different ways on~$A$ and~$B$: These are all~$2$-to-$1$-bit functions where~$(0,0)$ is mapped to~$0$ ($2^3$) minus the constant and minus the dependencies on~$A$ only and on~$B$ only.
In total, there are~$6\cdot 5=30$ such extremal points.
We are left with the last scenario:~$B$ depends on~$A$ and on~$C$, and~$C$ depends on~$A$ and on~$B$, {\it i.e.},~$A\preceq B$,~$C\preceq B$,~$A\preceq C$, and~$B\preceq C$. 
The constant, as above, is~$0$.
Given the random variable~$O_A$ fed to the environment by~$A$, the environment can {\em either\/} describe a channel from~$B$ to~$C$ ($B\preceq C$) {\em or\/} describe a channel from~$C$ to~$B$ ($C\preceq B$); any other channel would lead to a causal loop.
The direction of the channel must differ under different values~$o_A$ fed-back by~$A$, as otherwise~$B$ and~$C$ would not mutually depend on each other.
We have two possibilities on the direction given the value fed-back by~$A$ is~$O_A=0$.
Assume the direction to be~$B\preceq C$.
For the case~$O_A=0$, the channels from~$A$ to~$B$ and from~$B$ to~$C$ are the identity channels in order to comply with the fixed point~$(0,0,0)$.
In the other case~$O_A=1$, the direction of the channel between~$B$ and~$C$ is in the reverse direction compared to~$O_A=0$, {\it i.e.},~$C\preceq B$.
Then, because of~$O_A\not=0$, the random variables~$I_B$ and~$I_C$ are not forced to be~$(0,0)$; there exist two channels from~$A$ to~$C$ and another two channels from~$B$ to~$C$.
Therefore, we are left with~$3 \cdot 2\cdot 4=24$ possibilities.
An overview over these setups is given in Figure~\ref{fig:dep}.

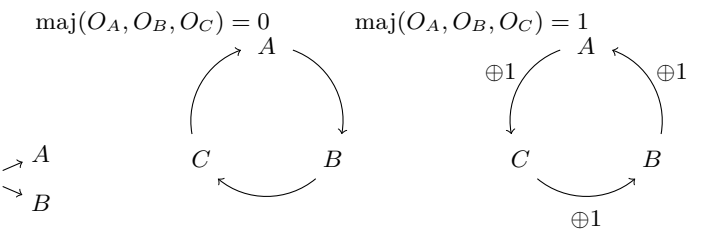
\begin{figure}
	\centering
	\begin{tikzpicture}
		\def\r{1}
		\def\d{20}
		\def\dx{2.1}
		\def\hd{.5}

		\draw[->] (-\dx,0)++(90-\d:\r) arc (90-\d:-30+\d:\r);
		\draw[->] (-\dx,0)++(-30-\d:\r) arc (-30-\d:-150+\d:\r);
		\draw[->] (-\dx,0)++(-150-\d:\r) arc (-150-\d:-270+\d:\r);
		\draw (-\dx,0)++(90:\r) node (A) {$A$};
		\draw (-\dx,0)++(-30:\r) node (B) {$B$};
		\draw (-\dx,0)++(-150:\r) node (C) {$C$};

		\draw[<-] (\dx,0)++(90-\d:\r) arc (90-\d:-30+\d:\r);
		\draw[<-] (\dx,0)++(-30-\d:\r) arc (-30-\d:-150+\d:\r);
		\draw[<-] (\dx,0)++(-150-\d:\r) arc (-150-\d:-270+\d:\r);
		\draw (\dx,0)++(90:\r) node (A2) {$A$};
		\draw (\dx,0)++(-30:\r) node (B2) {$B$};
		\draw (\dx,0)++(-150:\r) node (C2) {$C$};
		\draw (\dx,0)++(90-60:\r+.3) node (n1) {$\oplus 1$};
		\draw (\dx,0)++(-30-60:\r+.3) node (n2) {$\oplus 1$};
		\draw (\dx,0)++(-150-60:\r+.3) node (n3) {$\oplus 1$};

		\draw (-\dx-\r-\hd,\r+0.3) node (half1) {$\mathrm{maj}(O_A,O_B,O_C)=0$};
		\draw (\dx-\r-\hd,\r+0.3) node (half2) {$\mathrm{maj}(O_A,O_B,O_C)=1$};
	\end{tikzpicture}
	\caption{The left channel is chosen if the majority of the values fed into the environment is~$0$, otherwise, the right channel is chosen (see Equation~\ref{eq:game2}).}
	\label{fig:examples}
\end{figure}

The last setup (see Figure~\ref{fig:3to3}) where {\em no party\/} receives a constant builds a family of~$8$ extremal points.
All extremal points are equivalent up to relabelling of the inputs to and outputs from the environment.
One such extremal point is
\begin{align}
	\hat E_{\text{det}1}=
	\begin{pmatrix}
		1&0&0&0&0&0&0&1\\
		0&0&1&1&0&0&0&0\\
		0&0&0&0&1&0&1&0\\
		0&0&0&0&0&0&0&0\\
		0&1&0&0&0&1&0&0\\
		0&0&0&0&0&0&0&0\\
		0&0&0&0&0&0&0&0\\
		0&0&0&0&0&0&0&0
	\end{pmatrix}
	\,.
\end{align}
The behavior of this extremal point is
\begin{align}
	I_A=\bar O_BO_C\,,\quad I_B=O_A\bar O_C\,,\quad I_C=\bar O_AO_B
	\label{eq:game2}
	\,,
\end{align}
where~$\bar x$ is the negation of~$x$.
This solution is depicted in Figure~\ref{fig:examples}.
Its location in the polytope is shown in Figure~\ref{fig:slice}.

\section{Causal games}
Given a deterministic extremal point where every party is in the causal past of all other parties, a causal game can be constructed that can be won perfectly in the framework presented here, but that is lost if one assumes a global time.
\begin{definition}[Causal game]
	\rm
	Let a deterministic process map~\mbox{$O=(O_0,O_1,\dots)$} to~\mbox{$I^O=(I^O_0,I^O_1,\dots)$}, where the~\mbox{$i$-th} entry belongs to party~$S_i$, and where for every~$i$,~$I^O_i$ depends on all other parties' inputs to the environment.
We define a {\em causal game\/} where party~$S_i$ gets a random~$A_i$ and has to produce~$X_i=I^A_i$.
The parties are allowed to communicate in a predefined causal order~\cite{Baumeler:2013wy,Baumeler:2014cw,Oreshkov:2015vs}.
Let the guesses of all parties be~\mbox{$X=(X_0,X_1,\dots)$}, and let the random inputs to all parties be~\mbox{$A=(A_0,A_1,\dots)$}.
In a setup with~$n$ parties and where every party obtains and sends a~$d$-dimensional state, the game's winning probability is
\begin{align}
	p_\text{succ}=\frac{1}{d^n}\sum_I\Pr(X=I^A\,|\,A)\label{eq:succprobab}
	\,.
\end{align}
\end{definition}

Let~$p_\text{succ}^\text{C}$,~$p_\text{succ}^\text{NC}$ be the success probability of the game~\eqref{eq:succprobab} with, without the assumption of a predefined causal order, respectively.

\begin{theorem}[No winning strategy with predefined order]
	Using a predefined causal order, the success probability~\eqref{eq:succprobab} is strictly less than~$1$, {\it i.e.},~$p_\text{succ}^\text{C}<1$.
\end{theorem}
{\proof
For every party~$S_i$, the random variable~$I^A_i$ party~$S_i$ has to guess {\em depends\/} on the other parties' inputs~$A_{j(\not=i)}$.
In a predefined causal order, however, at least one party {\em is not in the causal future of any other party}.
Let~$S_i$ be that party, {\it i.e.},~\mbox{$\forall j\not=i:S_i\not\succeq S_j$} (see Lemma~\ref{def:predef}).
Then, at least for one input~$A_i=a'$ to~$S_i$, the party~$S_i$ cannot guess perfectly.
The success probability~$p_\text{succ}^\text{C}$ is upper bounded by
\begin{align}
	p_\text{succ}^\text{C}&= \frac{1}{d^n}\Bigg(\sum_{A,A_i\not=a'}\Pr(X=I^A\,|\,A)\\
	&\quad+\sum_{A,A_i=a'}\Pr(X=I^A\,|\,A) \Bigg)\\
	&\leq\frac{1}{d^n}\Bigg((d-1)d^{n-1}+ \sum_{A,A_i=a'}\Pr(X=I^A\,|\,A) \Bigg)
	\,.
\end{align}
The guessing probability for the non-perfect guess is upper bounded by
\begin{align}
	\Pr(X=I^A\,|\,A,A_i=a') \leq \frac{d^{n-1}-1}{d^{n-1}}
\end{align}
because for at least one input~$A$ with~\mbox{$A_i=a'$}, party~$S_i$ guesses wrongly.
Therefore, we obtain
\begin{align}
	p_\text{succ}^\text{C}&\leq \frac{1}{d^n}\left( (d-1)d^{n-1}+d^{n-1}\frac{d^{n-1}-1}{d^{n-1}} \right)\\
	&=\frac{1}{d^n}\left( d^n-d^{n-1}+d^{n-1}-1 \right)\\
	&=1-\frac{1}{d^n}
	\,.
\end{align}
\qed}

\begin{theorem}[Winning strategy without predefined causal order]
	If we drop the assumption of a predefined causal order, then the causal game~\eqref{eq:succprobab} can be won perfectly, {\it i.e.},~$p_\text{succ}^\text{NC}=1$.
\end{theorem}
{\proof
	To win the game perfectly, the parties use the process that maps the random variable~\mbox{$O=(O_0,O_1,\dots)$} to the random variable~\mbox{$I^O=(I^O_0,I^O_1,\dots)$} deterministically, forward their random input to the environment~\mbox{$(A_i=O_i)$}, and use the value obtained from the environment as their guess~$(X_i=I^O_i)$.
\qed}

For other games, a larger gap between the success probability with a predefined causal order and the success probability without a predefined causal order can be achieved --- as is shown in the examples below.

\section{Examples}
We briefly discuss two examples in the three-party scenario.
Let~$A$,~$B$,~$C$ be random input bits to the three parties~$A$,~$B$,~$C$, respectively, and let~$X$,~$Y$,~$Z$ be the corresponding output bits.
\begin{example}
	\rm
An extremal point of the first class of polytopes for three parties and binary inputs/outputs is
\begin{align}
	\hat E_\text{ex1}=\frac{1}{2}
	\begin{pmatrix}
		1&0&0&0&0&0&0&1\\
		0&0&1&0&0&1&0&0\\
		0&0&0&1&1&0&0&0\\
		0&1&0&0&0&0&1&0\\
		0&1&0&0&0&0&1&0\\
		0&0&0&1&1&0&0&0\\
		0&0&1&0&0&1&0&0\\
		1&0&0&0&0&0&0&1
	\end{pmatrix}
	\,;
\end{align}
its behavior is shown in Figure~\ref{fig:exampleprobab}.
This extremal point is a {\em proper mixture\/} of logically inconsistent processes, as it cannot be written as a convex combination of deterministic points from within the polytope; the left and the right channels from Figure~\ref{fig:exampleprobab} individually describe a causal loop and, hence, are logically inconsistent.
Initially, this process was used to show that in the classical scenario with three parties or more, correlations without predefined causal order can arise~\cite{Baumeler:2014cw}.
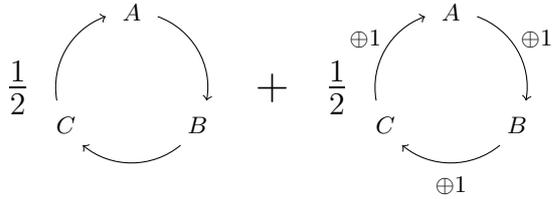
\begin{figure}
        \centering
        \begin{tikzpicture}
                \def\r{1}
                \def\d{20}
                \def\dx{2.1}
                \def\hd{.5}

                \draw[->] (-\dx,0)++(90-\d:\r) arc (90-\d:-30+\d:\r);
                \draw[->] (-\dx,0)++(-30-\d:\r) arc (-30-\d:-150+\d:\r);
                \draw[->] (-\dx,0)++(-150-\d:\r) arc (-150-\d:-270+\d:\r);
                \draw (-\dx,0)++(90:\r) node (A) {$A$};
                \draw (-\dx,0)++(-30:\r) node (B) {$B$};
                \draw (-\dx,0)++(-150:\r) node (C) {$C$};

                \draw[->] (\dx,0)++(90-\d:\r) arc (90-\d:-30+\d:\r);
                \draw[->] (\dx,0)++(-30-\d:\r) arc (-30-\d:-150+\d:\r);
                \draw[->] (\dx,0)++(-150-\d:\r) arc (-150-\d:-270+\d:\r);
                \draw (\dx,0)++(90:\r) node (A2) {$A$};
                \draw (\dx,0)++(-30:\r) node (B2) {$B$};
                \draw (\dx,0)++(-150:\r) node (C2) {$C$};
                \draw (\dx,0)++(90-60:\r+.3) node (n1) {$\oplus 1$};
                \draw (\dx,0)++(-30-60:\r+.3) node (n2) {$\oplus 1$};
                \draw (\dx,0)++(-150-60:\r+.3) node (n3) {$\oplus 1$};

                \draw (-\hd/2,0) node (PLUS) {\LARGE $+$};
                \draw (-\dx-\r-\hd,0) node (half1) {\LARGE $\frac{1}{2}$};
                \draw (\dx-\r-\hd,0) node (half2) {\LARGE $\frac{1}{2}$};
        \end{tikzpicture}
	\caption{Channel where the circular identity channel is uniformly mixed with the circular bit-flip channel.}
	\label{fig:exampleprobab}
\end{figure}
A causal game that can be formulated for this extremal point is
\begin{align}
	p_\text{succ}^\text{ex1}=\frac{1}{3}( \Pr&(X=B\oplus C\,|\,m=1)\\
	+\Pr&(Y=A\oplus C\,|\,m=2)\\
	+\Pr&(Z=A\oplus B\,|\,m=3))
	\,,
\end{align}
where, depending on the shared random trit~$m$, the party selected by~$m$ has to guess the parity of the other two parties' inputs.
If we assume a predefined causal order, then this causal game can be won with probability at most~$5/6$~\cite{Baumeler:2013wy}.
The reason for this is that at least one party is not in the causal future of the others.
This party, hence, can guess the parity with a probability of~$1/2$ only.
However, by using the environment from Figure~\ref{fig:exampleprobab}, one can win the game perfectly.
To achieve this, if~\mbox{$m=1$}, then party~$B$ forwards the random input~$(O_B=B)$, party~$C$ forwards the parity of the random input and the random variable obtained from the environment~$(O_C=C\oplus I_C)$, and party~$A$ uses the random variable obtained from the environment as its guess~$(X=I_A)$.
For the cases~$m=2$ and~$m=3$, the same strategy is used, but where the parties are permuted.
\end{example}

\begin{example}
	\label{ex:2}
	\rm
Another example~\cite{MateusAdrien} is depicted in Figure~\ref{fig:examples} and is a deterministic extremal point of the polytope with three parties and binary inputs/outputs (see also Equation~\eqref{eq:game2}).
Consider the causal game
\begin{align}
	p_\text{succ}^\text{ex2}=\frac{1}{2}(&\Pr(X=C,Y=A,Z=B\,|\,\mathrm{maj}(A,B,C)=0)\notag\\
	+\Pr&(X=\bar B,Y=\bar C,Z=\bar A\,|\,\mathrm{maj}(A,B,C)=1))
	\,,\notag
\end{align}
where~$\mathrm{maj}(A,B,C)$ is the majority of the three bits~$A$,~$B$, and~$C$.
Whenever the majority of the inputs is~0, {\it i.e.},~$\mathrm{maj}(A,B,C)=0$, then the parties play the ``guess-your-neighbours-input'' game~\cite{Almeida:2010fq,Acin:2012wi}: Party~$A$ guesses the input of party~$B$, party~$B$ guesses the input of party~$C$, and finally party~$C$ guesses the input of party~$A$;
the game is won if all guesses are correct simultaneously.
If the majority of the inputs is~1, then they play the same game in reverse direction and flip the output bits.
The success probability of winning this game in a world with a predefined causal order is upper bounded by~$3/4$.
This can be seen by the following reasoning.
In a predefined causal order, at least a single party has to make a guess without learning anything from the other parties.
For example, if party~$A$ causally precedes~$B$ and~$C$, {\it i.e.},~($A\preceq B$ and~$A\preceq C$), then party~$A$ at best always outputs~$0$ (see Table~\ref{tab:ex2}).
\begin{table}
	\centering
	\begin{tabular}{ccc|c|ccc}
		$A$&$B$&$C$&$\mathrm{maj}(A,B,C)$&$X$&$Y$&$Z$\\
		\hline
		0&0&0&0&0&0&0\\
		0&0&1&0&1&0&0\\
		0&1&0&0&0&0&1\\
		0&1&1&1&0&0&1\\
		1&0&0&0&0&1&0\\
		1&0&1&1&1&0&0\\
		1&1&0&1&0&1&0\\
		1&1&1&1&0&0&0
	\end{tabular}
	\caption{Conditions for winning the game of Example~\ref{ex:2}.}
	\label{tab:ex2}
\end{table}
By making such a guess, however, in~2 out of~8 cases, the parties will loose the game, yielding an upper bound of~$3/4$ to the success probability.
The same upper bound is achieved by choosing party~$B$ or party~$C$ as causally preceding the others.

By using the environment shown in Figure~\ref{fig:examples}, the game can be won perfectly.
The parties simply forward their inputs to the environment and use the bits obtained from the environment as the guesses.
\end{example}

%\begin{example}
%	\label{ex:3}
%	\rm
%\begin{table}
%	\centering
%	\begin{tabular}{ccc|c|ccc}
%		$A$&$B$&$C$&$f(A,B,C)$&$X$&$Y$&$Z$\\
%		\hline
%		0&0&0&1&0&0&0\\
%		0&0&1&0&0&0&0\\
%		0&1&0&1&0&0&1\\
%		0&1&1&1&1&0&1\\
%		1&0&0&0&0&1&1\\
%		1&0&1&0&0&0&1\\
%		1&1&0&1&0&1&1\\
%		1&1&1&0&1&0&1
%	\end{tabular}
%	\caption{Conditions for winning the game of Example~\ref{ex:3}.}
%	\label{tab:ex3}
%\end{table}
%The last example is depicted in Figure~\ref{fig:examplelast} and is deduced from the deterministic extremal point of the polytope with three parties and binary inputs/outputs given by Equation~\eqref{eq:game1}.
%The causal game
%\begin{align}
%	p_\text{succ}^\text{ex3}=\frac{1}{2}(
%	&\Pr(X=C,Y=A,Z=B\,|\,f(A,B,C)=1)\notag\\
%	+\Pr&(X=B,Y=\bar C,Z=A\,|\,f(A,B,C)=0))\notag
%	\,,
%\end{align}
%with
%\begin{align}
%	f(A,B,C)=\bar A\bar C\vee B(A\oplus C)
%	\,,
%\end{align}
%can be won perfectly by using the environment from Figure~\ref{fig:examplelast}.
%However, when we assume a predefined causal order, then the game cannot be won with a higher probability than~$3/4$.
%By looking at Table~\ref{tab:ex3} containing the conditions to win the game and the value of~$f(A,B,C)$, we can apply the same reasoning as for Example~\ref{ex:2}.
%
%Again, by using the process given by Equation~\eqref{eq:game1}, and where the parties use the strategy of forwarding the inputs to the environment, and then again use the states obtained from the environment as guesses, the game is won perfectly.
%\end{example}

\section{Conclusion and open questions}
We describe the polytope for classical multi-party processes without predefined causal order but where the arising correlations are logically consistent.
We also describe the polytope formed by {\em deterministic\/} extremal points; it excludes processes that are {\em proper mixtures\/} of logically inconsistent processes, {\it i.e.},~processes that cannot be written as a convex combination of deterministic processes from {\em within\/} the polytope.
For three parties or more, these polytopes contain processes that cannot be simulated by using a predefined causal order among the parties --- this is shown by violations of so-called causal inequalities.

A representation with polytopes helps for finding new causal games as well as for optimizing the processes for winning causal games; the optimization problem can be stated as a linear program.

In comparison, it has been shown that the set of {\em causal\/} correlations, {\it i.e.}, correlations {\em with\/} predefined causal order, also forms a polytope~\cite{Oreshkov:2015vs,araujo14b}.
A complete characterization for the two-party case is given~\cite{araujo14b}, however, in the multi-party case, a characterization is missing.
Such a characterization is interesting as then one could subtract it from the polytope studied in this work; this yields an exact characterization of the {\em non-causal\/} processes.
Another open question is to decide for which causal games the quantum correlations without causal order outperform their classical counterpart.

\begin{acknowledgments}
	We thank Mateus Ara{\'u}jo, Veronika Baumann, Cyril Branciard, {\v C}aslav Brukner, Fabio Costa, Adrien Feix, Arne Hansen, Alberto Montina, and Benno Salwey for helpful discussions.
	Furthermore we thank the anonymous referees for the detailed comments.
	The present work was supported by the Swiss National Science Foundation (SNF) and the National Centre of Competence in Research ``Quantum Science and Technology'' (QSIT).
\end{acknowledgments}

\bibliography{refsarxiv}

\end{document}